\documentclass{article}
\usepackage{hyperref}
\usepackage{amsmath,amssymb,booktabs,cleveref,setspace}
\usepackage{graphicx,subcaption,bm,authblk,comment}
\usepackage[left=1in, right=1in, top=1in, bottom=1in]{geometry}

\title{Sharp quadrature error bounds for the nearest-neighbor discretization of the regularized stokeslet boundary integral equation}

\author{Meurig T. Gallagher$^1$, Debajyoti Choudhuri$^2$, David J. Smith$^3$}

\affil{m.t.gallagher@bham.ac.uk$^{1}$, d.j.smith@bham.ac.uk$^3$, School of Mathematics, University of Birmingham,Birmingham. B15 2TT. UK.	choudhurid@nitrkl.ac.in$^2$, Department of Mathematics, National Institute of Technology, Rourkela, Odisha, India. 769 008.}

\date{}

\begin{document}

\maketitle

\begin{abstract}
	The method of regularized stokeslets is a powerful numerical method to solve the Stokes flow equations for problems in biological fluid mechanics. A recent variation of this method incorporates a nearest-neighbor discretization to improve accuracy and efficiency while maintaining the ease-of-implementation of the original meshless method. This method contains three sources of numerical error, the regularization error associated from using the regularized form of the boundary integral equations (with parameter $\varepsilon$), and two sources of discretization error associated with the force and quadrature discretizations (with lengthscales $h_f$ and $h_q$). A key issue to address is the quadrature error: initial work has not fully explained observed numerical convergence phenomena. In the present manuscript we construct sharp quadrature error bounds for the nearest-neighbor discretisation, noting that the error for a single evaluation of the kernel depends on the smallest distance ($\delta$) between these discretization sets. The quadrature error bounds are described for two cases: with disjoint sets ($\delta>0$)  being close to linear in $h_q$ and insensitive to $\varepsilon$, and contained sets ($\delta=0$) being quadratic in $h_q$ with inverse dependence on $\varepsilon$. The practical implications of these error bounds are discussed with reference to the condition number of the matrix system for the nearest-neighbor method, with the analysis revealing that the condition number is insensitive to $\varepsilon$ for disjoint sets, and grows linearly with $\varepsilon$ for contained sets. Error bounds for the general case ($\delta\geq 0$) are revealed to be proportional to the sum of the errors for each case. 
\end{abstract}

\section{Introduction}\label{sec:intro}

The development of numerical methods for the solution of Stokes flow has had significant impact on the study of problems in biological fluid dynamics \cite{klinteberg2016,pozrikidis1992,pozrikidis2002,veerapaneni2009,veerapaneni2011,zinchenko2000} and vice-versa. While there have been many powerful methods developed over the past few decades, one of the most effective and accessible tools for solving such problems is the method of regularized stokeslets, conceived of and developed by Cortez and colleagues \cite{ainley2008,cortez2001,cortez2005,cortez2014,cortez2012,cortez2015}, and recently extended to incorporate the use of the fast multipole method \cite{rostami2016}. A key advantage of this method over previous offerings is the meshless nature of the implementation, saving the significant investment of time and effort it takes to generate a mesh (particularly when dealing with complex biomolecular or cellular structures), potentially assisting with automation for applications in image analysis. These methods have had significant impact on a wide-ranging set of applications; a Google Scholar search on 25\textsuperscript{th} May 2018 with the term ``regularized stokeslets'' yielded $147$ results over the past year alone.

A new variation on the method of regularized stokeslets was recently proposed by Smith \cite{smith2018}, who uses a nearest-neighbor discretization of the regularized stokeslet boundary integral equation to improve the accuracy and efficiency of the classic Nystr\"om discretization \cite{cortez2005} while retaining the advantages of a meshless method. The computational efficiency of this new method, together with its extension to problems of locomotion in Stokes flow \cite{gallagher2018}, enables the study of previously computationally intractable problems such as improving the detailed modelling of the embryonic node of mice and zebrafish to incorporate Brownian and other effects, thus improving physiological accuracy. 

Mathematical details of the nearest-neighbor discretization will be provided in \cref{sec:mathintro}, for now we note that the sources of error for the method are threefold: the regularization error associated with using a regularized form of the boundary integral equations (with regularization parameter $ \varepsilon $), and two sources of discretization error associated with the approximating the the integral of the kernel at quadrature points with fine discretization lengthscale $ h_q $, and with approximating the forces with coarser discretization lengthscale $h_f$. While the original paper of Smith \cite{smith2018} provided an initial estimate of the error for the nearest-neighbor method in terms of these two discretization lengthscales and regularization parameter, these error bounds were noted not to be sharp as they did not fully replicate the sensitivity to $ \varepsilon $ suggested by the analysis (and seen in the classic Nystr\"om discretization). In the present work we not only provide the detailed analysis for calculating sharp error bounds for the nearest-neighbor discretization (see \cref{sec:analysis}), in doing this we uncover why the original work did not see the dependence on the regularization parameter $ \varepsilon $ and detail the situations where this dependence exists. We then also consider how these errors scale up in solving a practical problems (\cref{sec:practical}). Each of these analyses are then confirmed with numerical experiments (in \cref{sec:numExp}). These results will provide clear guidance for the best choices of discretization lengthscales and regularization parameter for given computational scenarios.

\subsection{Mathematical background}
\label{sec:mathintro}

The dimensionless form of the Stokes flow equations, which describe the very low Reynolds number fluid dynamics associated with sperm and cilia, is given by
\begin{equation}
-\bm{\nabla}p+\nabla^2\bm{u}=0, \quad \nabla \cdot \bm{u}=0,\label{eq:stokes}
\end{equation}
augmented with the no-slip, no-penetration boundary condition 
\(\bm{u}(\bm{X})=\dot{\bm{X}}\) for boundary points \(\bm{X}\), where overdot denotes time-derivative.

Regularized stokeslet methods involve representing the flow field around a body \(B\) by an integral of the form,
\begin{equation}
u_j(\bm{x})=-\frac{1}{8\pi}\iint_B S_{jk}^\varepsilon(\bm{x},\bm{y})f_k(\bm{y})dS_{\bm{y}}, \label{eq:bie}
\end{equation}
where \(S_{jk}^\varepsilon\) is the velocity part of the solution to the Stokes flow equations \cref{eq:stokes} driven by a smoothed point force in the $ k- $direction, with regularization parameter \(\varepsilon>0\). The most widely-studied example \cite{cortez2005} is for 3D flow and takes the form,
\begin{equation}
  S_{ij}^\varepsilon(\bm{x},\bm{y}) = \delta_{ij}\frac{|\bm{x}-\bm{y}|^2+2\varepsilon^2}{(|\bm{x}-\bm{y}|^{2}+\varepsilon^2)^{3/2}}+\frac{(x_i-y_i)(x_j-y_j)}{(|\bm{x}-\bm{y}|^{2}+\varepsilon^2)^{3/2}}.
  \label{eqn:stokelets}
\end{equation}
The limiting form of this kernel is the classical stokeslet or Oseen tensor,
\begin{equation}
S_{jk}(\bm{x},\bm{y}) = 
\frac{\delta_{jk}}{|\bm{x}-\bm{y}|}+\frac{(x_j-y_j)(x_k-y_k)}{|\bm{x}-\bm{y}|^3}. 
\end{equation}

Regularized stokeslet methods are implemented numerically by imposing equation~\cref{eq:bie} for \(\bm{x}=\bm{X}\in B\) together with the condition \(\bm{u}(\bm{X})=\dot{\bm{X}}\) (collocation), followed by discretization of the unknown traction \(\bm{f}(\bm{y})\) and the numerical quadrature. The original (Nystr\"{o}m) discretization of \cref{eq:bie}, by Cortez \textit{et al.} \cite{cortez2005}, takes the form,
\begin{equation}
  u_j(\bm{X}[m])=\frac{1}{8\pi}\sum_{q=1}^Q S_{jk}^\varepsilon(\bm{X}[m],\bm{X}[q])F_k[q], \label{eq:nystrom}
\end{equation}
where \(\{\bm{X}[1],\ldots,\bm{X}[Q]\in B\}\) is a set of quadrature points, and the discretized force at \(\bm{X}[q]\) is written as $F_k[q] = -f_k(\bm{X}[q]) dS(\bm{X}[q])$. This method has the major advantage of implementational simplicity, a property which has resulted in widespread adoption; caveats are that the regularization parameter \(\varepsilon\) and discretization size \(h\) must be chosen in proportion, and that the typical size of the linear system (\(3Q\times 3Q\)) may be rather larger than would be required by a classical boundary integral method to achieve converged results. This scaling then limits the applications of the method when considering large or complex problems.

The constant-panel boundary element discretization suggested by Smith \textit{et al.} \cite{smith2009} takes the form,
\begin{equation}
  u_j(\bm{x}[m])=\frac{1}{8\pi}\sum_{n=1}^N f_k[n] \iint_{B_n} S_{jk}^\varepsilon(\bm{x}[m],\bm{y}) dS_{\bm{y}}, \label{eq:bem}
\end{equation}
where \(\{B_1,\ldots,B_N\}\) is a partitioning of the surface \(B\) (mesh) with centroids \({\bm{x}[n]\in B_n}\), and the discretization of the traction on \(B_n\) is denoted \(f_k[n]\). Because the near-field of the regularized stokeslet is rapidly-varying (resembling the function \((r^2+\varepsilon^2)^{-1/2}\) as \(r\rightarrow 0\)), this traction discretization need not be as refined as the quadrature discretization in equation~\cref{eq:nystrom}, i.e.\ one can take \(N\ll Q\). The stokeslet integral in equation~\cref{eq:bem} is still evaluated numerically via quadrature, however in contrast to equation~\cref{eq:nystrom}, the quadrature discretization does not affect the number of degrees of freedom of the resulting linear system. The boundary element discretization is therefore more efficient and accurate than the Nystr\"{o}m method, however it has the disadvantage of requiring true mesh generation rather than a simple list of surface points.

To attempt to combine the implementational simplicity of the Nystr\"{o}m method with the efficiency and accuracy of the boundary element discretization, Smith \cite{smith2018} proposed the use of a meshless nearest-neighbor method. Two discretizations are generated, a `coarse force' set \(\mathcal{F}=\{\bm{x}[1],\ldots ,\bm{x}[N]\}\) and a `fine quadrature' set \(\mathcal{Q}=\{\bm{X}[1],\ldots,\bm{X}[Q]\}\), with \(N<Q\). The force at the quadrature points \(\bm{f}(\bm{y})dS(\bm{X}[q])\) is approximated by its value at the nearest force point via a nearest-neighbor projection \(\bm{f}(\bm{X}[q])dS(\bm{X}[q])\approx \sum_{q=1}^Q \nu[q,n] \bm{f}(\bm{x}[n])dS(\bm{x}[n])\), where \(\nu[q,n]\) is a binary matrix with precisely one \(1\) in each row. The resulting linear system is then,
\begin{equation}
  u_j(\bm{x}[m])=\frac{1}{8\pi}\sum_{n=1}^N \left(\sum_{q=1}^Q S_{jk}^\varepsilon(\bm{x}[m],\bm{X}[q]) \nu[q,n] \right)F_k[n] , \label{eq:nndisc}
\end{equation}
where \(\bm{F}[n]= \bm{f}(\bm{x}[n])dS(\bm{x}[n])\).

Smith \cite{smith2018} conducted an initial analysis of the error associated with the nearest-neighbor method \cref{eq:nndisc}. In addition to a regularization error \(O(\varepsilon)\) found by Cortez \textit{et al.} \cite{cortez2005}, the error associated with discretization of the traction is \(O(h_f)\) (where \(h_f\) characterises the fineness of the force points), and the error associated with numerical quadrature was estimated as \(O(\varepsilon^{-2} h_f^2 h_q)+O(h_f^{-1} h_q)\) (where \(h_q\) characterises the fineness of the quadrature points; formal definitions in \cref{eq:hf} and \cref{eq:hq} below). It was noted that this error bound was not sharp because numerical experiments suggested that the error does not diverge for very small values of \(\varepsilon\), indeed the choice of $h_f$ as the lengthscale for quadrature discretisation error was somewhat arbitrary. In the present manuscript we will address this issue further. It will be shown that the quadrature error for a single evaluation of the kernel depends on the shortest distance from the force discretization ($\mathcal{F}$) to the quadrature discretization ($\mathcal{Q}$), denoted by $ \delta $. The error of the full problem is then discussed in terms of three distinct cases (detailed in \cref{fig:schematic}): i) when $ \delta > 0 $, i.e. the force and quadrature sets are disjoint, ii) when $ \delta = 0 $, i.e. every force point is also a quadrature point, and iii) when the force and quadrature sets are non-disjoint, but $ \delta = 0 $ for some points.

\section{Analysis of the quadrature error for a single kernel evaluation}
\label{sec:analysis}

The principal challenge regarding numerical quadrature concerns evaluation where the kernel is rapidly-varying, i.e.\ where \(|\bm{x}-\bm{y}|\) is `small'. For the Nystr\"{o}m discretization the near-field part of the integral is primarily evaluated in the sum~\cref{eq:nystrom} when \(q=m\). It is clear that this evaluation is problematic as \(\varepsilon\rightarrow 0\) because \(S_{jk}^\varepsilon(\bm{x},\bm{x})\rightarrow \varepsilon^{-1}\) as \(\varepsilon\rightarrow 0\). This divergence also underlies the \(O(\varepsilon^{-2}h_f^2h_q)\) term in the nearest-neighbor error estimate. A key advantage of the nearest-neighbor method however is that this situation can be avoided by ensuring that the force and quadrature discretizations are disjoint. We will denote the minimum distance between the discretizations by,
\begin{equation}
\delta = \min_{q=1,\ldots,Q} \min_{n=1,\ldots,N} |\bm{x}[n]-\bm{X}[q]|.\label{eq:mindist}
\end{equation}
If any overlap \(\bm{x}[n]=\bm{X}[q]\) occurs, then clearly \(\delta =0\).
We also recall from \cite{smith2018} the definitions characterising the fineness of the force and quadrature discretizations,
\begin{align}
h_f & = \max_{m=1,\ldots,N} \min_{n=1,\ldots,N; n\not = m} |\bm{x}[m]-\bm{x}[n]|,\label{eq:hf} \\
h_q & = \max_{p=1,\ldots,Q} \min_{q=1,\ldots,Q; q\not = p} |\bm{X}[p]-\bm{X}[q]|.\label{eq:hq}
\end{align}

\begin{figure}
	\centering
	\includegraphics[width=\textwidth]{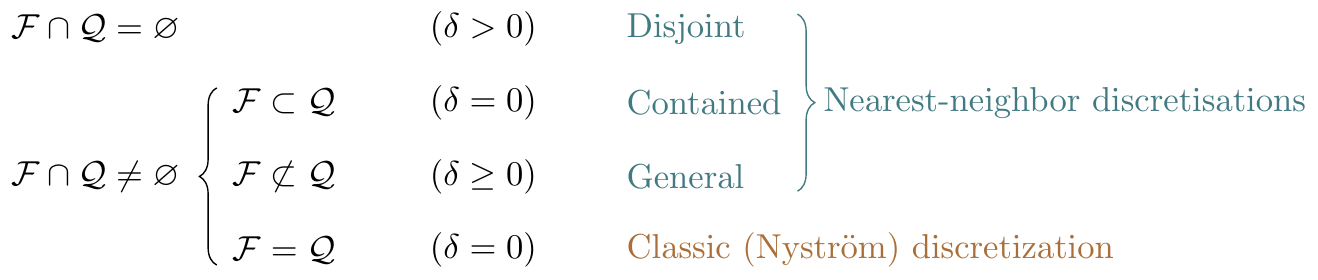}
	\caption{Schematic detailing the characterisations for potential force ($\mathcal{F}$) and quadrature ($\mathcal{Q}$) discretization sets with $ \delta $, denoting the minimum distance between the discretizations, defined in equation \cref{eq:mindist}. Each choice is further shown as belonging to either the set of nearest-neighbor or classic (Nystr\"om) discretizations.}
	\label{fig:schematic}
\end{figure}

We note then, that we can characterise the choices of force ($\mathcal{F}$) and quadrature ($\mathcal{Q}$) discretizations for the nearest-neighbor method as one of three possibilities: 

	\begin{enumerate}
		\item \textit{Disjoint}: $\mathcal{F}\cap \mathcal{Q} = \varnothing$ $\left(\delta > 0\right)$;
		\item \textit{Contained}: $\mathcal{F}\subset\mathcal{Q}$ $\left(\delta = 0\right)$;
		\item \textit{General}: $\mathcal{F}\cap\mathcal{Q}\neq\varnothing$ with $\mathcal{F}\not\subset\mathcal{Q}$ $\left(\delta \geq 0\right)$.
	\end{enumerate}
These cases are illustrated in the schematic provided in \cref{fig:schematic}, with the classic (Nystr\"om) discretization (which has $\delta = 0$) included for comparison. 

We will develop detailed analysis of the quadrature error of, in order, the disjoint case (\cref{sec:disjoint}), and the contained case (\cref{sec:contained}). Each analysis will be based on analysis of the error of approximation of \(\int_0^{2\pi} \int_0^1 K(r)rdrd\theta\) where \(K(r)=(r^2+\varepsilon^2)^{-1/2}\), which captures the near-singular behaviour of the kernel \(S_{jk}^\varepsilon(\bm{x},\bm{y})\) for small \(|\bm{x}-\bm{y}|\). In \cref{sec:practical} we will discuss how the quadrature errors scale for practical problems and discuss the general case of mixed disjoint and contained quadrature sets.

\subsection{Previous analysis}\label{sec:previous}
As discussed by Smith \cite{smith2018}, an estimate of quadrature error can be made using the mean value inequality \(|K(r)-K(0)|\leqslant M_1 r\), where \(M_1\) is a bound on \(|K'(r)|\). Because \(K'(r)=-r(r^2+\varepsilon^2)^{-3/2}\), we have for all constants \(a>\varepsilon\),
\begin{enumerate}
\item In the region \(0\leqslant r \leqslant a\), the bound \(M_1=O(\varepsilon^{-2})\).
\item In the region \(a\leqslant r\), the bound \(M_1=|K'(a)|\).
\end{enumerate}
Smith \cite{smith2018} used the above to split the quadrature into three regions, (i) \(0<r<h_f\), (ii) \(h_f<r<h_f^{1/2}\) and (iii) \(h_f^{1/2}<r<1\). The quadrature errors can be estimated from the values of \(M_1\), the area of the region, and the quadrature spacing. The resulting error estimates, for each region in turn, are then, (i) \(O(\varepsilon^{-2} h_f^2 h_q)\), (ii) \(O(h_f^{-1} h_q)\), (iii) \(O(h_f^{-1}h_q)\). 

\subsection{The disjoint case}\label{sec:disjoint}
To improve on this analysis we first address the case for which \(\delta>0\). This entails that there is an inner region \(0\leqslant r<\delta\) which contains no quadrature points, i.e.\ the region is neglected from the numerical quadrature. The error associated with this neglect can be calculated as,
\begin{equation}
L(\delta,\varepsilon)=\int_0^{2\pi} \int_0^1 K(r)rdrd\theta=2\pi\left[(\delta^2+\varepsilon^2)^{1/2}-\varepsilon\right]=O(\delta).
\end{equation}
This error estimate is valid provided \(0<\delta, \varepsilon \ll 1\) regardless of the relative sizes of \(\varepsilon\) and \(\delta\).

The remaining error can be calculated by a similar approach to \cref{sec:previous}. To achieve a sharp error estimate, we will consider a sequence of annuli \(\delta \leqslant r < h_q^{\phi_1}\), \(h_q^{\phi_1}\leqslant r < h_q^{\phi_2}, \ldots\), where \(\phi_1=1\) and \(\phi_1>\phi_2>\ldots\) (this analysis deals with the case \(\delta < h_q\); if \(\delta \geqslant h_q\) the error is no worse). The quadrature error for the first annulus is \(\delta^{-2} h_q^{2\phi_1+1}\) and for the \(n\)th annulus is \(O(h_q^{2(\phi_n-\phi_{n-1})+1})\). 

It is clear therefore that taking \(\phi_n-\phi_{n-1}\) to be small and negative will yield a close-to-optimal error estimate. For example, for any fixed integer \(P>3\) we may take \(\phi_n=1-(n-1)/P\) for \(n=1,\ldots,P+1\). The error for the first annulus is \(O(\delta^{-2}h_q^{3})=O((h_q/\delta)^2 h_q)\), and for the remaining annuli is \(O(h_q^{1-2/P})\). The total error over \(P\) annuli is therefore \(O((h_q/\delta)^2 h_q)+O(P h_q^{1-2/P})\). By taking increasingly large values of \(P\) the latter term approaches linear convergence. Therefore provided $ h_q/\delta = O\left(1\right) $, quadrature convergence is linear in $ h_q $ and insensitive to $ \varepsilon $.

In summary, the total error estimate (including regularization error, force discretization error, and quadrature error) for the nearest-neighbor regularized stokeslet method with disjoint discretizations is,
\begin{equation}
E_1 = O(\varepsilon)+O(h_f)+O((h_q/\delta)^2 h_q) + O(P h_q^{1-2/P}), \label{eq:nonoverlap}
\end{equation}
for any integer \(P>3\), where \(\delta\) is the minimum distance from the force discretization to the quadrature discretization, as defined in equation~\cref{eq:mindist}.

\subsection{The contained case}\label{sec:contained}
The analysis of \cref{sec:previous} is based on three regions parameterised by \(h_f\). The argument is in fact valid with \(h_f\) replaced by any lengthscale \(\lambda > \varepsilon/\sqrt{2}\) so that the local maximum of \(|K'(r)|\) appears inside the inner circle \(0\leqslant r < \lambda\). By similar arguments to the above (based on taking annuli of radius \(\lambda^{1-(n-1)/P}\)) we then have the total error estimate,
\begin{equation}
O(\varepsilon)+O(h_f)+O(\varepsilon^{-2}\lambda^2 h_q)+O(P\lambda^{-2/P}h_q). \label{eq:overlap}
\end{equation}
Taking \(\lambda=\varepsilon^{1/2}h_q^{1/2}\), yields an error,
\begin{equation}
E_2 = O(\varepsilon)+O(h_f)+O(\varepsilon^{-1}h_q^2)+O(P\varepsilon^{-1/P} h_q^{1-1/P}). \label{eq:overlap2}
\end{equation}
As \(\varepsilon\rightarrow 0\), the dominant term in the above is \(O(\varepsilon^{-1}h_q^2)\), which has the very advantageous property of being quadratic in \(h_q\), but an unwanted inverse dependence on \(\varepsilon\). 
It is therefore clear for the contained case that we cannot expect to be able to reduce \(\varepsilon\) independently of \(h_q\). 

\begin{figure}[t]
	\centering
	\begin{subfigure}[t]{\textwidth}
		\includegraphics[width=\textwidth]{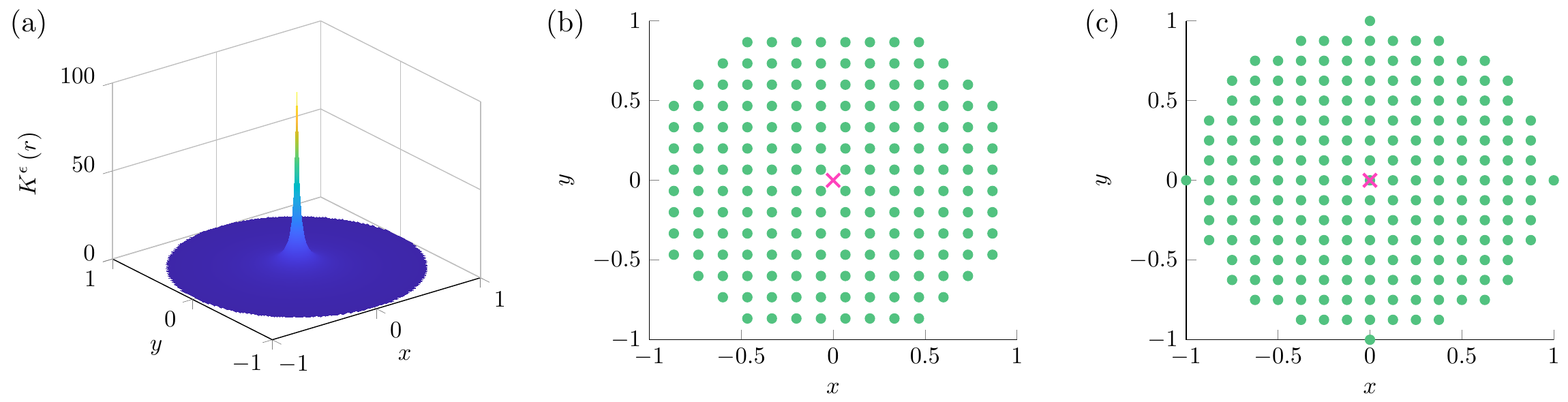}
		\phantomcaption
		\label{fig:kernel_a}
	\end{subfigure}
	\begin{subfigure}[t]{0\textwidth}
		\phantomcaption
		\label{fig:kernel_b}
	\end{subfigure}
	\begin{subfigure}[t]{0\textwidth}
		\phantomcaption
		\label{fig:kernel_c}
	\end{subfigure}
	\caption{Set up for numerical quadrature experiments. (a) The near-singular kernel \(K^\varepsilon(r)\) plotted for \(r<1\), with \(\varepsilon=0.01\). (b) Depiction of a disjoint quadrature set with \(h_q=1.33\). (c) Depiction of an contained quadrature set with \(h_q=0.125\).}
	\label{fig:kernel}
\end{figure}

\section{Practical implications of the quadrature error}
\label{sec:practical}

Having developed the analysis to understand the quadrature error inherent in a single evaluation of the kernel, it is of practical use to assess how this error scales in a full application of the nearest-neighbor discretization (when solving a resistance problem for example). When numerically constructing and solving the matrix system $ A\bm{\chi} = \bm{b} $ the relative error in the calculation of $ \bm{\chi} $ in terms of small deviations in the construction of matrix $ A $, $ \Delta A $, is bounded by
\begin{equation}
	{\Big{\|}}\frac{\Delta\bm{\chi}}{\bm{\chi}}{\Big{\|}} \leq {\big{\|}}A^{-1}{\big{\|}}{\big{\|}}A{\big{\|}}\frac{\big{\|}\Delta A\big{\|}}{\big{\|}A\big{\|}} = \mathrm{cond}\left(A\right)\frac{\big{\|}\Delta A\big{\|}}{\big{\|}A\big{\|}},
\end{equation}
where $ \mathrm{cond}\left(A\right) $ represents the condition number of $ A $, and ${{\|}\Delta A{\|}}/{{\|}}A{\|}$ is the relative error in the numerical construction of $ A $. The analysis of \cref{sec:analysis} provides the error estimates for the size of $\|\Delta A\|/\|A\|$ for both disjoint (\cref{sec:disjoint}) and contained (\cref{sec:contained}) quadrature sets; to build an understanding of the error in a practical application of the nearest-neighbor method it thus remains to understand how the condition number of the matrix $ A $ behaves.

\begin{figure}[t]
	\centering
	\begin{subfigure}[t]{\textwidth}
	\centering
		\includegraphics[width=0.8\textwidth]{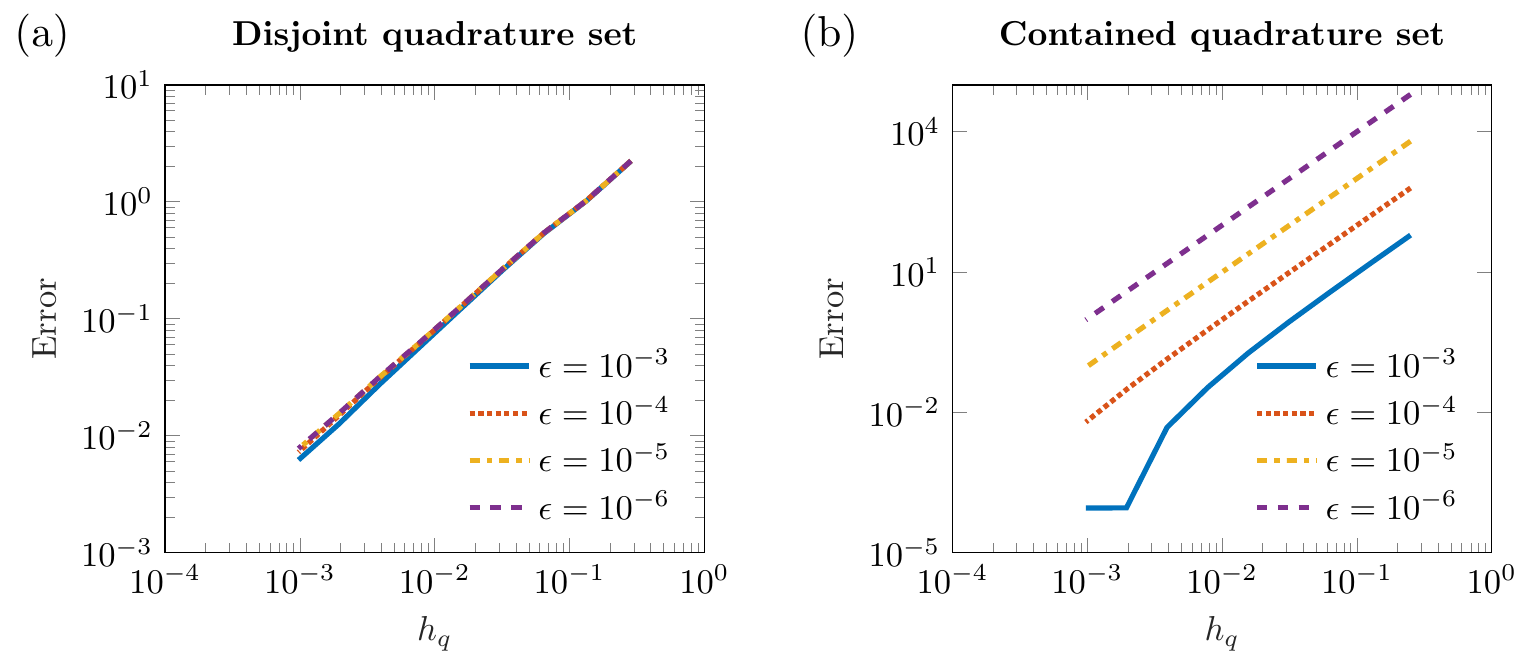}
		\phantomcaption
		\label{fig:quadratureConv_a}
	\end{subfigure}
	\begin{subfigure}[t]{0\textwidth}
		\phantomcaption
		\label{fig:quadratureConv_b}
	\end{subfigure}
	\caption{Convergence of numerical quadrature of the kernel \(K^\varepsilon\) with number of points for four values of the regularization parameter \(\varepsilon\). (a) Disjoint quadrature set, (b) Contained quadrature set.}
	\label{fig:quadratureConv}
\end{figure}

Each row of the matrix $ A $ consists of a diagonal entry which comes from evaluation of of $ S_{ij}^{\varepsilon} $ \cref{eqn:stokelets} at the force and associated nearest-neighbor quadrature points. Consequently, for numerically tractable numbers of quadrature points $ Q $, the diagonal entries of $ A $ have lower bound
\begin{equation}
	A_{ii}\geq \overline{A} \propto \begin{cases}
		\left(\delta^2 + \varepsilon^2\right)^{-1/2} & \text{for a disjoint quadrature set},\\
		\varepsilon^{-1} & \text{for a contained quadrature set}.		
	\end{cases}
	\label{eqn:diags}
\end{equation} 
In the case that $ Q $ becomes large then the change in the diagonal entries of $ A $ due to the evaluation of $ S_{ij}^{\varepsilon} $ at many points will become significant; we will explore this numerically in \cref{sec:numExp}, however for practical densities of quadrature points this source of error is insignificant compared to the dominant ($\varepsilon^{-1}$) term. Denoting the sum of the off-diagonal elements (corresponding to a surface integral over a fixed area) by $ C_Q $, which will grow with increasing numbers of quadrature points $ Q $, we can apply the Gershgorin circle theorem \cite{bell1965gershgorin} to show that all eigenvalues of $ A $ lie in a circle of radius $ C_Q $ about the diagonal values in \cref{eqn:diags}. The ratio between the largest and smallest eigenvalues (the condition number) is therefore bounded by
\begin{equation}
	\frac{1/\varepsilon + C_Q}{1/\varepsilon - C_Q}\sim 1 + 2 C_Q\varepsilon,
\end{equation} 
for a contained quadrature set, and, for a disjoint quadrature set, as $ \varepsilon \rightarrow 0 $, 
\begin{equation}
	\frac{1/\delta_{\mathrm{min}}+C_Q}{1/\delta_{\mathrm{max}}-C_Q} \sim \frac{\delta_{\mathrm{max}}}{\delta_{\mathrm{min}}}\left( 1 + C_Q\left(\delta_{\mathrm{min}} + \delta_{\mathrm{max}}\right)\right),
\end{equation}
where $ \delta_{\mathrm{min}} $ and $ \delta_{\mathrm{max}} $ are the smallest and largest of the distances $ \delta $ (as calculated in \cref{eq:mindist}) between each force and quadrature discretization sets, with $ \delta_\mathrm{max}\ll 1 $ (and assuming $ \delta_{\mathrm{max}}/\delta_{\mathrm{min}} = O\left(1\right) $). Provided again that $ Q $ is not too large (which will cause $ C_Q $ to correspondingly increase), is is clear that this method resolves a problem that affects boundary element methods for Stokes flow: ensuring that the condition number remains bounded as the size of the force elements approaches zero.

The bounds that this analysis places on the condition number of the matrix $A$ are practically very useful when solving problems with the nearest-neighbor discretization. There may be situations where it is desirable to discretize a subject with both disjoint and contained quadrature sets; when considering a biological swimmer, for example, it may be helpful to consider separately the discretization of flagellum and body. We can thus consider the error of the \textit{general case} as being composed of the error from the \textit{disjoint case}, $E_1$ \cref{eq:nonoverlap}, plus the error from the \textit{contained case}, $E_2$ \cref{eq:overlap2}, multiplied by the condition number of the matrix $ A $.

\section{Numerical experiments}\label{sec:numExp}
We will now confirm the analysis through numerical experiments. \Cref{sec:quadExp} will consider the convergence of numerical quadrature of the function \(K^\varepsilon(r)=(r^2+\varepsilon^2)^{-1/2}\) for a region including \(r=0\); \cref{sec:cond} will investigate the condition number of the matrix $ A $, and \cref{sec:resExp} the resistance tensor, each owing to the problem of a prolate spheroid undergoing rigid body motion.

\subsection{Quadrature convergence}\label{sec:quadExp}
The function \(K^\varepsilon(r)\) is illustrated in \cref{fig:kernel_a}. Two types of quadrature method are illustrated: disjoint quadrature (\cref{fig:kernel_b}) for which the quadrature set does not include the origin, and contained quadrature (\cref{fig:kernel_c}) for which the quadrature set does include the origin. Numerical results with these quadrature sets are shown in \cref{fig:quadratureConv}. The disjoint quadrature set (\cref{fig:quadratureConv_a}) exhibits approximately linear convergence with \(h_q\) and is insensitive to \(\varepsilon\), as expected from equation~\cref{eq:nonoverlap}. The contained quadrature set performs very well for \(\varepsilon=0.001\), exhibiting approximately quadratic convergence, as expected from equation~\cref{eq:overlap2}. However, as also predicted, the absolute error shows an approximate \(\varepsilon^{-1}\) dependence, becoming highly inaccurate for \(\varepsilon\leqslant 10^{-5}\).

\begin{figure}[tp]
	\centering
	\includegraphics[width=8cm]{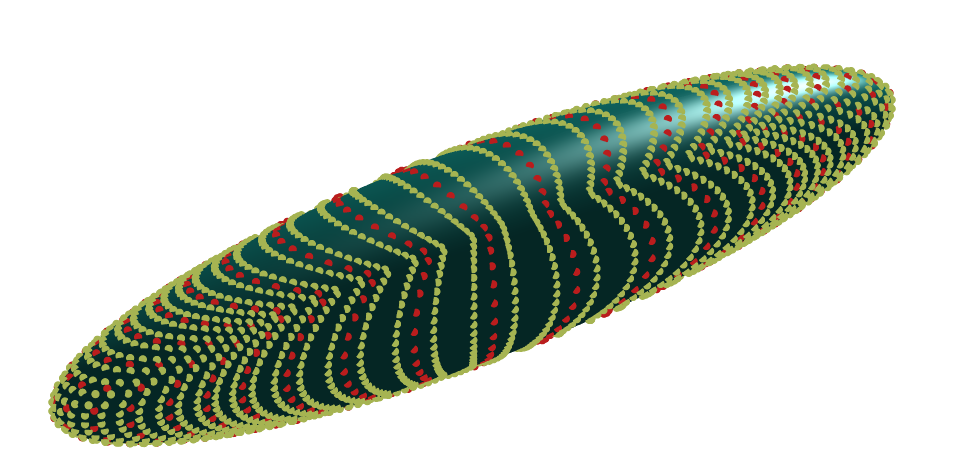}
	\caption{Sketch of the nearest-neighbor discretization of a prolate spheroid. Here, the red dots show the force discretization, with a disjoint quadrature set shown in light green.}
	\label{fig:prolate}
\end{figure}

\subsection{Condition number}\label{sec:cond}

\begin{figure}[tp]
	\centering
	\begin{subfigure}[t]{\textwidth}
		\includegraphics[width=\textwidth]{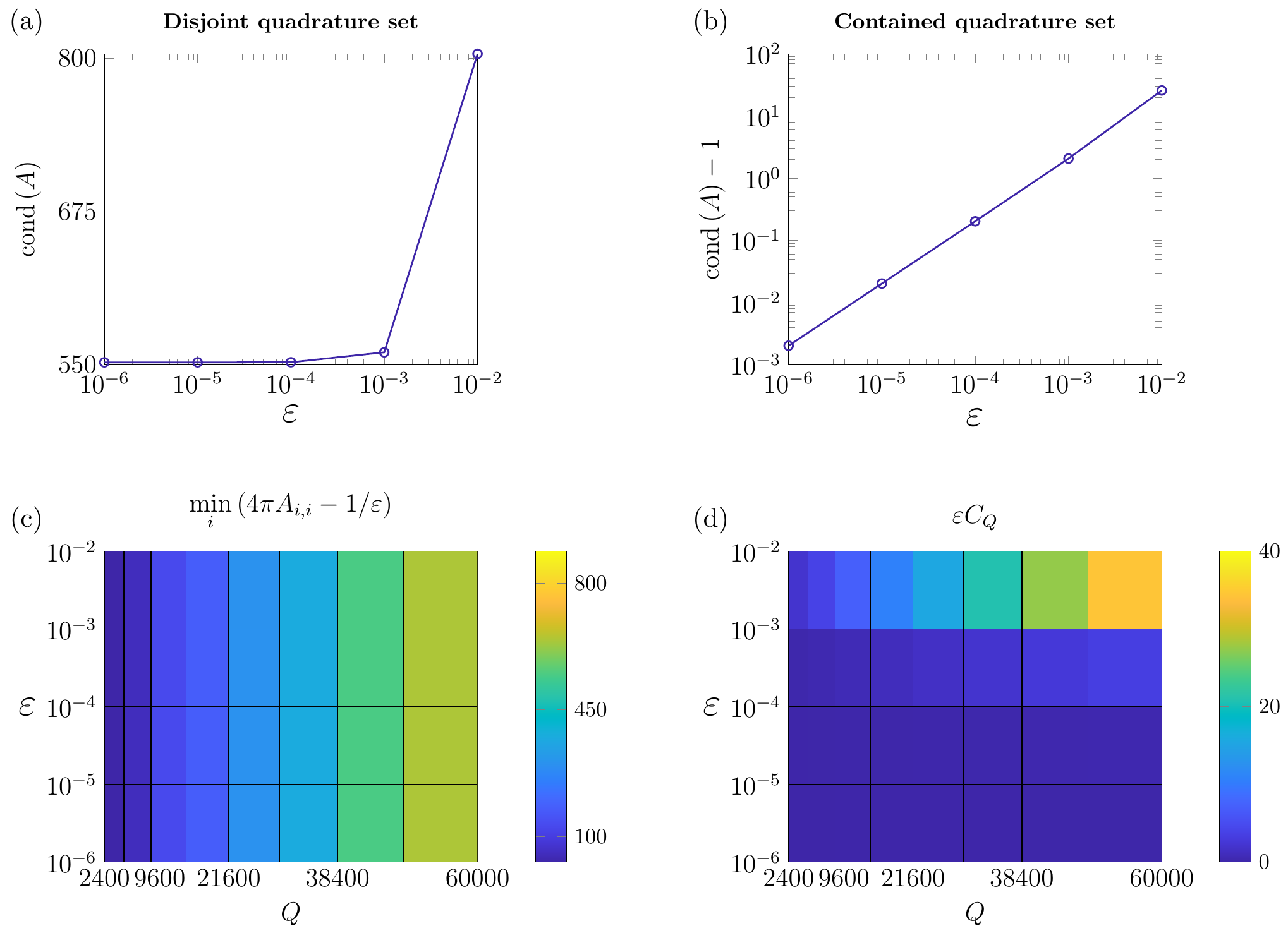}
		\phantomcaption
		\label{fig:condNoOverlapEps}
	\end{subfigure}
	\begin{subfigure}[t]{0\textwidth}
		\phantomcaption
		\label{fig:condOverlapEps}
	\end{subfigure}
	\begin{subfigure}[t]{0\textwidth}
		\phantomcaption
		\label{fig:condOverlapAii}
	\end{subfigure}
	\begin{subfigure}[t]{0\textwidth}
		\phantomcaption
		\label{fig:condOverlap_Cq}
	\end{subfigure}
	\caption{Analysis of the condition number and diagonal entries of the matrix $ A $. Panel (a) plots the condition number $\mathrm{cond}\left(A\right)$ against $ \varepsilon $ for a disjoint quadrature set, and shows the robustness of the condition number to changes as $ \varepsilon\rightarrow 0 $. Panel (b) plots $ \mathrm{cond}\left(A\right) - 1 $ against $ \varepsilon $ for a contained quadrature set, and shows approximate linear dependence with a slope of $ 1.02 $. Panel (c) shows the minimum deviation of the diagonal entries of $A$ away from $1/\varepsilon$ for large numbers of quadrature points $ Q $ with a contained quadrature set. Panel (d) shows the $ C_Q $, the maximum sum of the off-diagonal entries of $A$, multiplied by $ \varepsilon $ for large numbers of quadrature points $Q$ with a contained quadrature set.}
	\label{fig:conditionNo}
\end{figure}

To assess the condition number analysis of \cref{sec:practical} on a relevant problem we follow Smith \cite{smith2018} and construct the matrix $ A $ owing to the resistance problem of a prolate spheroid associated with rigid body motion. In \cref{fig:condNoOverlapEps} we plot the condition number of $ A $ against decreasing values of $ \varepsilon $ for a disjoint quadrature set and, as predicted, we see that the condition number plateaus rapidly as $ \varepsilon\rightarrow 0 $. For a contained quadrature set we plot in \cref{fig:condOverlapEps} $ \mathrm{cond}\left(A\right) - 1 $, against the same values of $ \varepsilon $, where we see the approximate linear dependence on this quantity with $ \varepsilon $ (the slope in the figure is calculated as approximately $ 1.02 $). To assess the predictions regarding the diagonal entries of the matrix $ A $, in \cref{fig:condOverlapAii} we plot the minimum increase in diagonal elements of $ A $ from $ 1/\epsilon $ against increasing numbers of quadrature points $ Q $; here we clearly see that the diagonal terms are indeed bounded with $ A_{ii} \geq 1/\varepsilon $. In \cref{fig:condOverlap_Cq} we plot the maximum row sum of off-diagonal entries $ C_Q $ multiplied by $ \varepsilon  $. Here we see that $ \varepsilon C_Q $ grows slowly with $ Q $, however for computationally practical values of $ Q $ the size of $ \varepsilon C_Q $ (and thus the condition number of $ A $) remain manageable. These numerical results agree with the analysis of \cref{sec:practical}, and give confidence to the calculated error bounds.

\subsection{Resistance problem}\label{sec:resExp}

\begin{figure}[tp]
	\centering
	\begin{subfigure}[t]{\textwidth}
		\centering
		\includegraphics[width=0.6\textwidth]{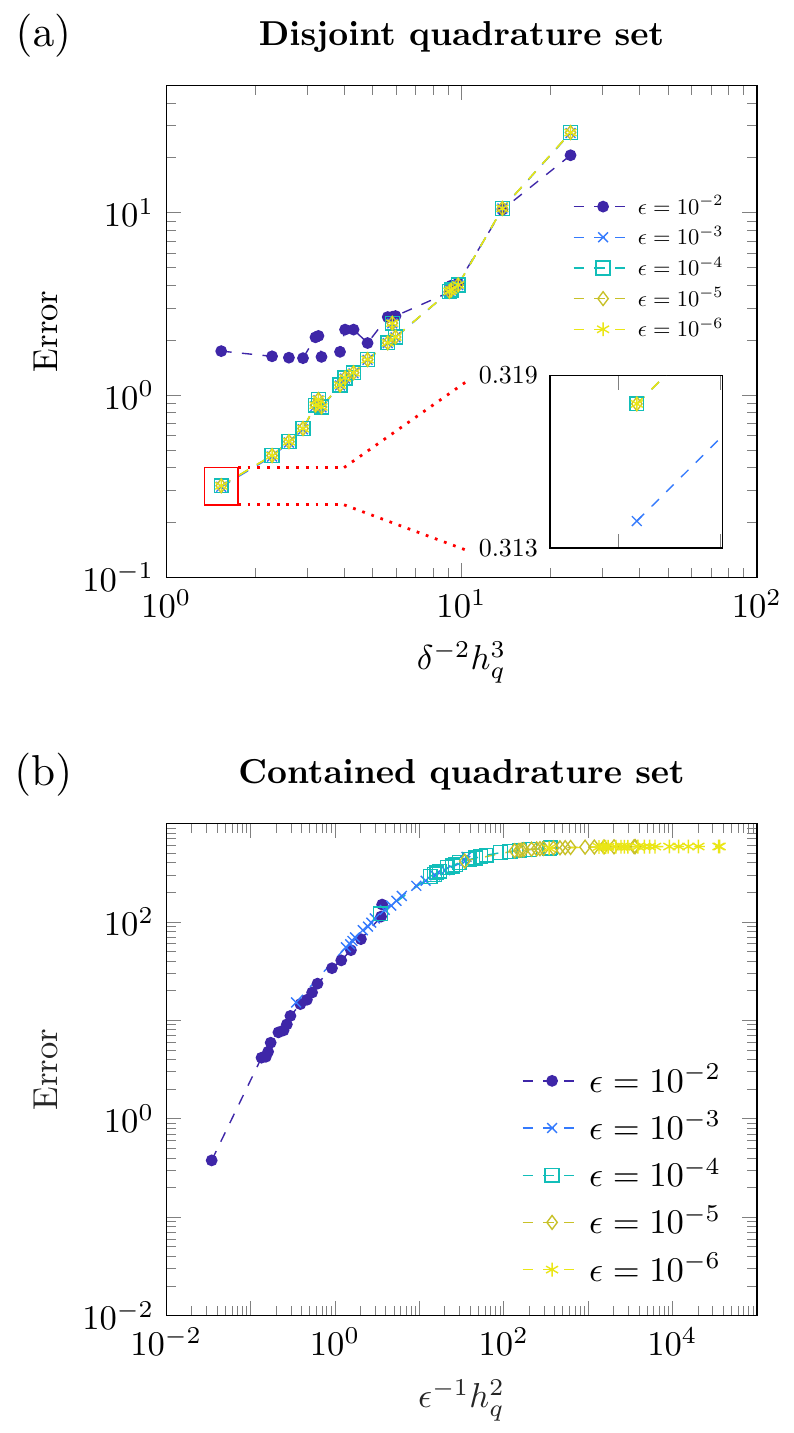}
		\phantomcaption
		\label{fig:resProbNoOverlap}
	\end{subfigure}
	\begin{subfigure}[t]{0\textwidth}
		\phantomcaption
		\label{fig:resProbOverlap}
	\end{subfigure}
	\caption{Absolute error in calculating the resistance tensor of a prolate spheroid undergoing rigid body motion for (a) disjoint and (b) contained quadrature sets. In (a) we see convergence with decreasing $\delta^{-2}h_q^3$ with an insert showing a zoomed view emphasising the lack of $ \varepsilon $ dependence for $ \varepsilon < 10^{-3} $. Panel (b) shows convergence with decreasing $\varepsilon^{-1} h_q^2$. Each panel has been calculated for five values of the regularization parameter $ \varepsilon $ with fixed $ h_f \approx 0.249 $.}	\label{fig:resProbError}
\end{figure}

To assess the analysis of \cref{sec:analysis} on a relevant problem we follow Smith \cite{smith2018} and calculate the resistance tensor of a prolate spheroid, with an axis ratio of $5$, associated with rigid body motion. This problem has the added benefit of an analytical solution with which to compare  \cite{kim2013}. A rendering of the force and quadrature discretizations for this problem is shown in \cref{fig:prolate}. We again test both disjoint and contained quadrature sets with results provided in \cref{fig:resProbError}. The disjoint case again shows approximately linear convergence with $ \delta^{-2}h_q^{3} $ and, if $ \varepsilon\leq 10^{-3} $, is very robust to the choice of $ \varepsilon $. This robustness in illustrated more clearly in \cref{tab:resProb} which contains a selection of the values used to plot \cref{fig:resProbNoOverlap}. The contained case exhibits near linear convergence in $\varepsilon^{-1}h_q^2$ for moderate values of $\epsilon$, and the error collapses onto a single curve. For this case we see, as predicted by the analysis in \cref{sec:analysis}, a clear dependence on $ \varepsilon $ with the relative error approaching $ 100\% $ for $ \varepsilon \leq 10^{-5} $. While we may naively expect the error to blow up for large $ \varepsilon^{-1}h_q^2 $, the limit of small $ \varepsilon $ for a contained quadrature set leads to a calculation of zero force, and thus a zero resistance tensor, over the prolate spheroid. This results in a relative error of $ 100\% $ in the limit $ \varepsilon\rightarrow 0 $.

\begin{table}[tp]
	\centering
	\caption{Error in calculating the resistance tensor of a prolate spheroid undergoing rigid body motion for a disjoint quadrature set (as depicted in \cref{fig:resProbNoOverlap}). The error is calculated for five choices of the regularization parameter $ \varepsilon $ and increasing $ \delta^{-2}h_q^3 $, with fixed $ h_f\approx 0.249 $.}
	\label{tab:resProb}	
	\begin{tabular}{c c c c c c c c}
	               & \multicolumn{7}{c}{$\delta^{-2}h_q^3$}\\
	    \cmidrule{2-8}           
		$\varepsilon$& $ 2.282$  & $3.355$    & $4.805$ & $5.975$ & $9.101$ & $13.75$ & $23.40$ \\
		\midrule
		$10^{-2}$ & $ 1.603$  & $ 1.619 $  & $1.929$ & $2.712$ & $3.666$  & $10.40$ & $20.66$   \\
		$10^{-3}$ & $ 0.4600$ & $ 0.8547 $ & $1.560$ & $2.072$ & $3.699$ &  $10.55$ & $27.35$  \\
		$10^{-4}$ & $ 0.4657$ & $ 0.8602 $ & $1.564$ & $2.081$ & $3.702$ &  $10.55$ & $27.52$  \\
		$10^{-5}$ & $ 0.4658$ & $ 0.8603 $ & $1.564$ & $2.081$ & $3.702$ &  $10.55$ & $27.52$  \\
		$10^{-6}$ & $ 0.4658$ & $ 0.8603 $ & $1.564$ & $2.081$ & $3.702$ &  $10.55$ & $27.53$  
	\end{tabular}
\end{table}

\section{Conclusions}

This paper has calculated sharp quadrature error bounds for the nearest-neighbor regularized stokeslet discretization. We have shown that this error depends on the shortest distance ($\delta$) from the force discretization ($\mathcal{F}$) to the quadrature discretization ($\mathcal{Q}$), and that the behaviour of the quadrature error can be characterised by two discrete cases. The total error in solving a Stokes flow problem using the nearest-neighbor discretization is either described by one of these cases, or by a general, mixed, case. We will now detail the characteristics of each of these cases in turn.

\subsection{The \textit{disjoint} case, $\mathcal{F}\cap\mathcal{Q} = \varnothing$ $ \left(\delta > 0\right) $}
	
	When the force and quadrature discretizations are \textit{disjoint}, the total quadrature error estimate for the nearest-neighbor regularized stokeslet method is
	\begin{equation*}
		E_1 = O\left(\varepsilon\right) + O\left({h_f}\right) + O\left({\delta^{-2}h_q^3}\right) + O\left({P h_q^{1-2/P}}\right),
	\end{equation*}
	for any integer constant $ P > 3 $, where $ \varepsilon \ll 1 $ is the stokeslet regularization parameter, and $ h_f $ and $ h_q $ are given by equations \cref{eq:hf} and \cref{eq:hq} respectively. In this case the quadrature error is very robust to the choice of $ \varepsilon $, with errors that are approximately linear in $h_q$ when $ \delta \sim h_q $.
			
	When solving practical problems (of the form $A\bm{\chi}=\bm{b}$) with a \textit{disjoint} quadrature set, the condition number of the matrix $A$ is also robust to the choice of $ \varepsilon $. This ensures that, providing $ Q $ is not too large, the error in solving a practical problem should remain $ E_1 $; this error result has been validated through solving the resistance problem of a prolate spheroid undergoing rigid body motion, displayed in \cref{fig:resProbNoOverlap}. 
			
\subsection{The \textit{contained} case, $\mathcal{F}\subset\mathcal{Q}$ $ \left(\delta = 0\right) $}
	
	When the force and quadrature discretizations are \textit{contained}, the total quadrature error estimate for the nearest-neighbor regularized stokeslet method is
	\begin{equation*}
		E_2 = O\left(\varepsilon\right) + O\left({h_f}\right) + O\left({\varepsilon^{-1}h_q^2}\right) + O\left({P \varepsilon^{-1/P} h_q^{1-1/P}}\right),
	\end{equation*}
	for any integer constant $ P > 3 $, where $ \varepsilon \ll 1 $ is the stokeslet regularization parameter, and $ h_f $ and $ h_q $ are given by equations \cref{eq:hf} and \cref{eq:hq} respectively. In this case the quadrature error is approximately quadratic in $h_q$, but also has an unwanted inverse dependence of $ \varepsilon $. It is clear that in this case we are not able to reduce $ \varepsilon $ independently of $h_q$.
			
	When solving practical problems 	(of the form $A\bm{\chi}=\bm{b}$) with a \textit{contained} quadrature set, the condition number of the matrix $ A $ grows linearly with $ \varepsilon $. This ensures that, providing $ Q $ is not too large, the error in solving a practical problem should remain $ E_2 $; this error result has been validated through solving the resistance problem of a prolate spheroid undergoing rigid body motion, displayed in \cref{fig:resProbOverlap}.

\subsection{The \textit{general} case, $\mathcal{F}\not\subset\mathcal{Q}$ ($\delta \geq 0$)}

	We have provided analyses for each of the \textit{disjoint} and \textit{contained} cases. It is conceivable that for practical purposes one may wish to use separate discretizations for different elements of a problem (for example discretizing a swimmer's body and flagella differently) inducing the combination of both disjoint and contained quadrature sets. The analysis of the condition number of the matrix system indicates that the total error for the solution of practical problems with general force and quadrature sets should be proportional to the sum $E_1 + E_2$. 

\subsection{Discussion}

The analyses were confirmed via numerical experiments: we have tested the convergence of numerical quadrature of the kernel $ K^{\varepsilon}\left(r\right) $ (\cref{sec:quadExp}), assessed the change in condition number of the matrix system (\cref{sec:cond}), and tested the calculation of the resistance tensor of a prolate spheroid undergoing rigid body motion (\cref{sec:resExp}). Each of these numerical experiments closely replicated the predictions of the analysis.

The analysis contained within the present work provides useful insight into the error inherent in using the nearest-neighbor discretization; we believe that the application of this analysis when choosing the parameters of the method for a given problem of interest will be valuable in ensuring that the desired convergence criteria are met.

The nearest-neighbor discretization For 

The approach of the nearest-neighbor method reduces degrees of freedom while retaining near field accuracy. For better scaling to large problems involving many far field evaluations it may be interesting to explore whether fast multipole implementations can be integrated into the nearest-neighbor method. If it is possible for such adaptations to be made while keeping the ease-of-implementation and simplicity of the present method then this will surely be valuable.

One area where the nearest-neighbor discretisation may prove useful is in the simulation of biological microswimmers: the meshless nature of this method leaves open the possibility for automated swimmer generation from the analysis of experimental imaging data, an option which is far from straightforward for methods which require the generation of a true mesh.

\section*{Acknowledgments}
This work was supported by Engineering and Physical Sciences Research Council award EP/N021096/1.

\bibliographystyle{siamplain}

\end{document}